\documentclass[11pt]{article}
\usepackage{amsmath}
\usepackage[margin=2.0cm]{geometry}
\usepackage{epsfig}
\usepackage{amsfonts}
\usepackage{titlesec}
\usepackage{fancyhdr}
\usepackage{setspace}
\usepackage{authblk}
\usepackage{subcaption}
\usepackage{xcolor}
\usepackage{cite}
\usepackage{tabularx}
\usepackage{epstopdf}
\usepackage{epsfig}
\usepackage{epstopdf-base}
\newcommand{\authorname}{HS Dhami \emph{et al}.} 
\titleformat{\section}
{\normalfont\bfseries\scshape}{\thesection}{1em}{}

\titleformat{\subsection}
{\normalfont\bfseries\scshape}{\thesubsection}{1em}{}

\titleformat{\subsubsection}
{\normalfont\slshape}{\thesubsubsection}{1em}{}

\titleformat{\title}{\normalfont\bfseries}{\thesection}{1em}{}

\title{\vspace{-3em} \large \bfseries Microstructure and Mechanical Properties of Heat-Treated High Nitrogen Martensitic Stainless Steel}
\author[1]{\normalsize Harish Singh Dhami \thanks{Email: harishdhami@iisc.ac.in}}
\author[2]{\normalsize Nellori Dileep Kumar }
\author[2]{\normalsize Thomas Tharian}
\author[3]{\normalsize Chakravarthy P}
\author[1]{\normalsize Koushik Viswanathan}
\affil[1]{\normalsize \textsl{Dept. of Mechanical Engineering, IISc, Bangalore}}
\affil[2]{\normalsize \textsl{Liquid Propulsion System Center(LPSC)/ISRO,Thiruvanathapuram}}
\affil[3]{\normalsize \textsl{Indian Institute of Space Science and Technology,Thiruvanathapuram}}
\date{\normalsize \textsc{\today}}

\begin{document}
\maketitle {}
\thispagestyle{plain}
\begin{abstract}
	The High Nitrogen Martensitic Stainless Steel (HNMS) was subjected to three different austenitizing cycles of 1050$^\circ$C, 1075$^\circ$C and 1100$^\circ$C followed by subzero treatment at -70$^\circ$C. The fraction of retained austenite has been reduced after sub-zero treatment as revealed by microstructural evolution. The material was subsequently tempered at different temperatures ranging from 180$^\circ$C to 650$^\circ$C and the change in micro-structure, hardness, tensile strength and toughness were investigated after each heat treatment cycle. Optical microscopy, electron microscopy with EDS and X-Ray diffraction techniques were used to characterize the material. This has showed the constituents of microstructure were lath martensite, precipitated metal carbides of type $M_{23}C_6$, $M_7C_6$ and carbo-nitrides. Hardness, tensile testing and Charpy impact testing were carried to evaluate mechanical properties after the heat treatment which has showed the better mechanical properties for the samples solutionised at 1075$^\circ$C. Secondary hardening has been observed on tempering above 450$^\circ$C which can be attributed to the precipitation of secondary phase inter-metallic compounds. Hardness attains a peak value at peculiar temperature range after which it decreases on further tempering which is most likely because of the loss of coherency of the precipitates with the metal matrix. This has been further confirmed by the XRD of the specimens before and after tempering. The study stablishes the structure-property correlation of HNMS for different heat treatment cycles. The results indicate that a good combination of hardness and strength can be achieved after solutionizing at 1075$^\circ$C followed by double tempering at 525$^\circ$C.
\end{abstract}
\section{Introduction}
Martensitic grade stainless steels are widely used in aerospace application owing to their superior properties at room as well at sub-zero temperatures \cite{isfahany2011effect,pant2013studies}. Bearings used in turbo pumps for cryogenic applications employ AISI 440C, a high carbon martensitic stainless steel which possesses the required properties such as high wear resistance, strength and moderate corrosion resistance \cite{girodin2002characterisation,fan2012effects}. Since carbon is the major constituent that imparts hardness as well as strength, and also leads to higher amount of retained austenite as well as the formation of coarser carbides \cite{gingell1997carbide}. However, the precipitation of coarse carbide segregation along the grain boundaries results in poor toughness and thus leads to poor corrosion resistance and therefore limits the usage of this material in many strategic applications\cite{hucklenbroich1999high,nykiel2014transformations}. It is very pertinent to have the combination of superior strength and toughness to avoid catastrophic and premature failures in service. To overcome the above said limitations, Nitrogen was thought of a substitute to replace a partial amount of carbon to strengthen the martensitic stainless steel \cite{byrnes1987nitrogen}. High nitrogen martensitic stainless (HNMS) steel has been developed, lowering the carbon content and adding Nitrogen in place to get the similar solid solution strengthening\cite{speidel1992high,hucklenbroich1999high, ojima2008origin}. This work attempts to study the properties such as toughness, strength and corrosion resistance by optimizing the heat treatment parameters.

\section{Materials and Methodology}
The material was procured in the form of rods of diameter 55mm. The chemical composition analysis as determined by Atomic Absorption spectroscopy and XRF, is furnished in Table\ref{table:1}.

The as received rods were reduced to 20 mm through hot rolling process, which were sliced into bars of 200mm length for conducting heat treatment trials. The samples were subjected to Annealing treatment in which they were heated to 850$^\circ$C for 360minutes followed by furnace cooling. The annealed samples were subsequently Hardened by heating it to various austenitizing temperatures such as 1050$^\circ$C, 1075$^\circ$C and 1100$^\circ$C for soaking time of 30minutes in vacuum furnace followed by Argon gas quenching at 1bar. In order to reduce the amount of retained austenite the samples were subjected to sub-zero treatment at -70 $^\circ$C for 120 minutes. Immediately after the hardening treatment, samples were tempered at temperatures of 150, 180, 250, 350, 450, 525, 550 and 650$^\circ$C for 60 minutes. The hardness was measured using Rockwell hardness testing machine with 160kg load applied for 10 second swell time. The samples tempered at 525$^\circ$C were further tempered at the same temperature (double tempering) in order to obtain better mechanical properties. X-ray diffraction (XRD) was performed using Cr K$\alpha$ radiation (1.5403\AA) for identification of different phases that could have formed during heat treatment. For the purpose of obtaining microstructure, all the samples were polished to mirror finish using standard emery papers and etched using 15$ml$ $HCl$ mixed with 7$ml$ $HNO_3$ along with few drops of glycerol. The etched samples were viewed using optical microscope and scanning electron microscope to obtain the details of microstructure. The samples were subjected to tensile test using INSTRON with a velocity of 2mm/min with 8MPa free load. The load and displacement was continuously monitored using a data acquisition system.
\begin{table}[ht]
	\caption{Elemental composition of HNMS Steel} 
	\centering 
	\begin{tabular}{c c c c c c c c} 
		\hline 
		C & Si & Mn & Cr & Mo & V & N & Fe\\ [0.5ex] 
		\hline 
		 0.45&0.42&0.30&17.0&1.70&0.20&0.17& Bal.\\ 

		\hline 
	\end{tabular}
	\label{table:1} 
\end{table}

\section{Results and Discussion}

\subsection{Microstructural Observations}
Optical microscopic characterization were used to analyze the microstructural development and identify phase distribution after each austenitizing and subsequent tempering treatment. 
\begin{figure*}[h!]
	\centering
	\includegraphics[width=0.6\textwidth]{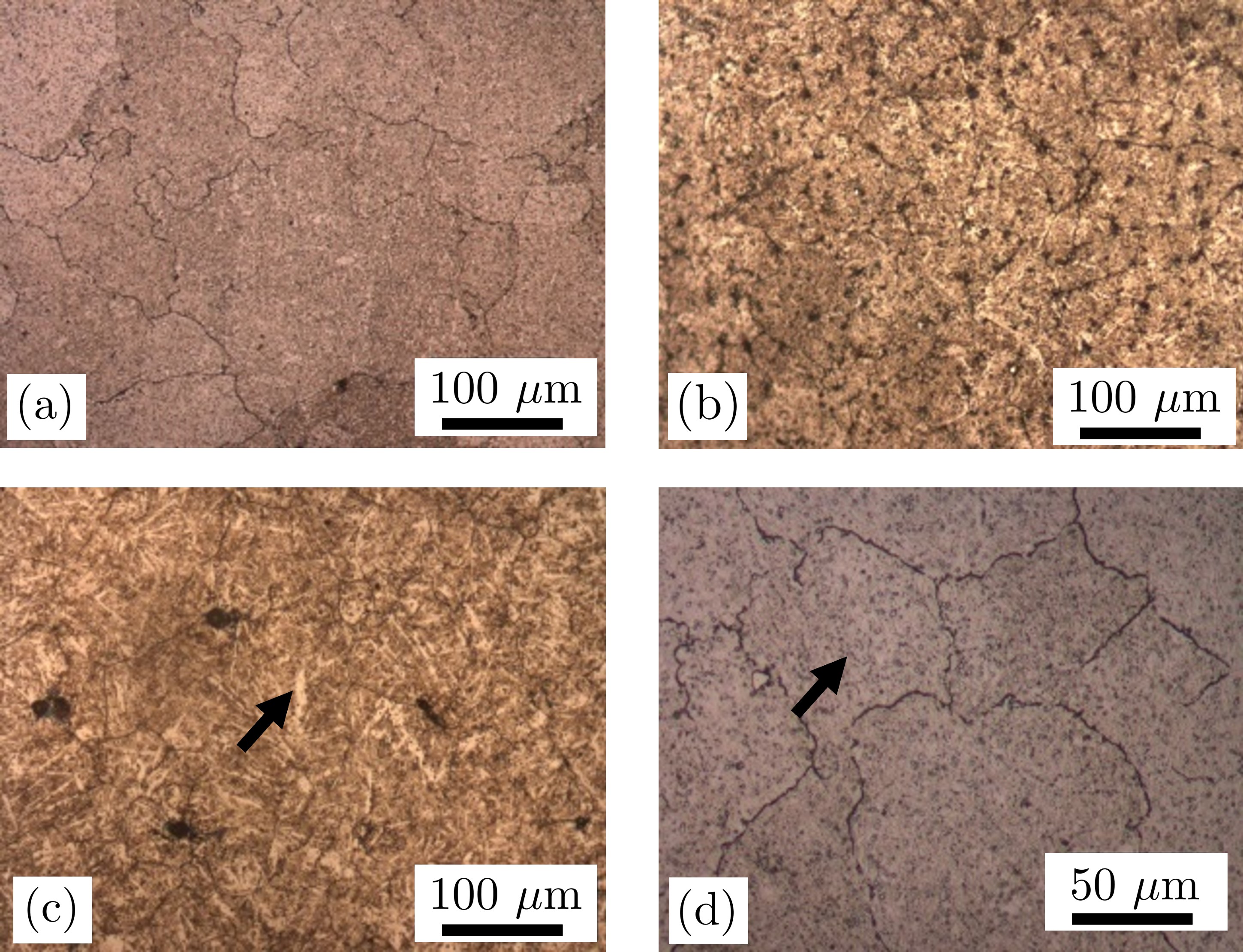}
	\caption{(a)(b)(c) Optical microstructure 200$\times$ after hardening at 1050$^\circ$C,1075$^\circ$C and 1100$^\circ$C respectively (arrow showing retained austenite) (d) 1050$^\circ$C at500$\times$ (without tempering)}
	\label{micro_1}
\end{figure*}

Fig.\ref{micro_1} (a) (b) (c) shows the optical micrographs of HNMS steel hardened at different austenitizing temperatures. It can be clearly observed that the fraction of retained austenite increased on austenitizing at 1100$^\circ$C compared to the 1050$^\circ$C. Fig.\ref{micro_1} (d) shows the morphology and distribution of carbides (spherical) in martensitic matrix at higher magnification. 

\begin{figure*}[h!]
	\centering
	\includegraphics[width=0.6\textwidth]{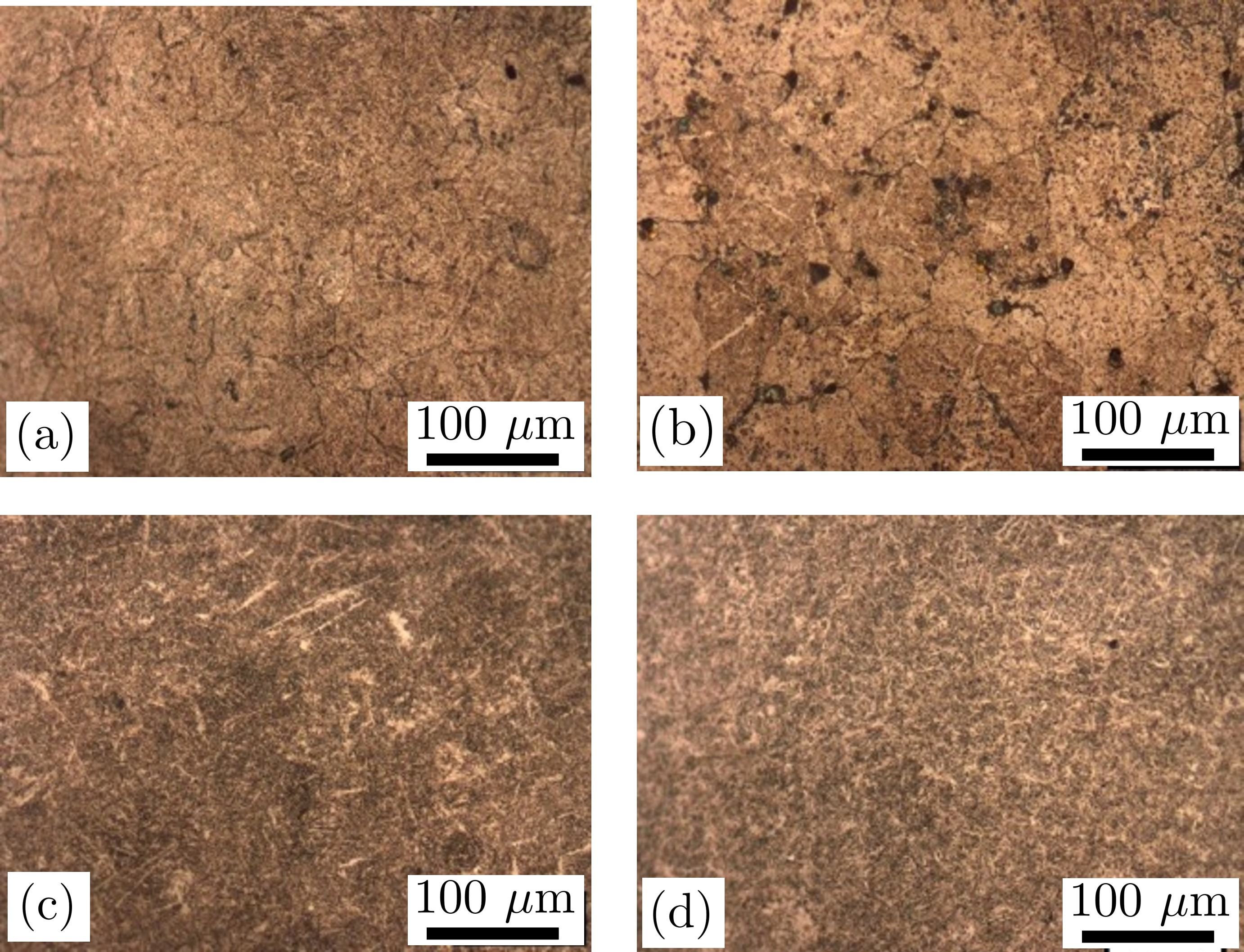}
	\caption{(a) (b) (c) (d) Optical Micrographs at 200$\times$ after Austenitized at 1075° C followed by tempering at 180$^\circ$C, 350$^\circ$C, 525$^\circ$C and 525$^\circ$C (double tempered) respectively }
	\label{micro_2}
\end{figure*}
Fig.\ref{micro_2} (a) (b) (c) (d) shows optical micrographs of HNMS steel hardened at 1075$^\circ$C, followed by tempering at 180, 350, 525 and 525$^\circ$C double tempered respectively. The microstructure mainly comprises of lath martensite and carbides which gets darkened during etching indicating the tempered martensitic structure. Selective etching was performed to reveal the carbides.

\begin{figure*}[h!]
	\centering
	\includegraphics[width=0.6\textwidth]{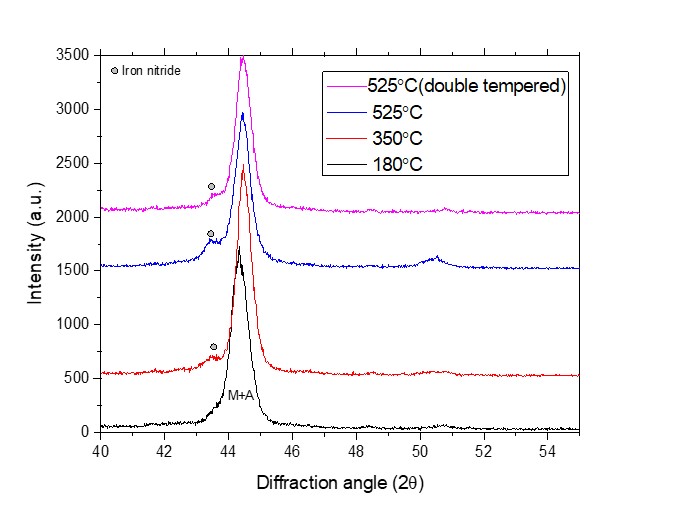}
	\caption{X-Ray diffraction (XRD) spectrum after austenitizing at 1075$^\circ$C and subsequent tempering. Precipitate formation can be seen as shoulder peaks (circle)}
	\label{XRD}
\end{figure*}
\begin{figure*}[h!]
	\centering
	\includegraphics[width=0.6\textwidth]{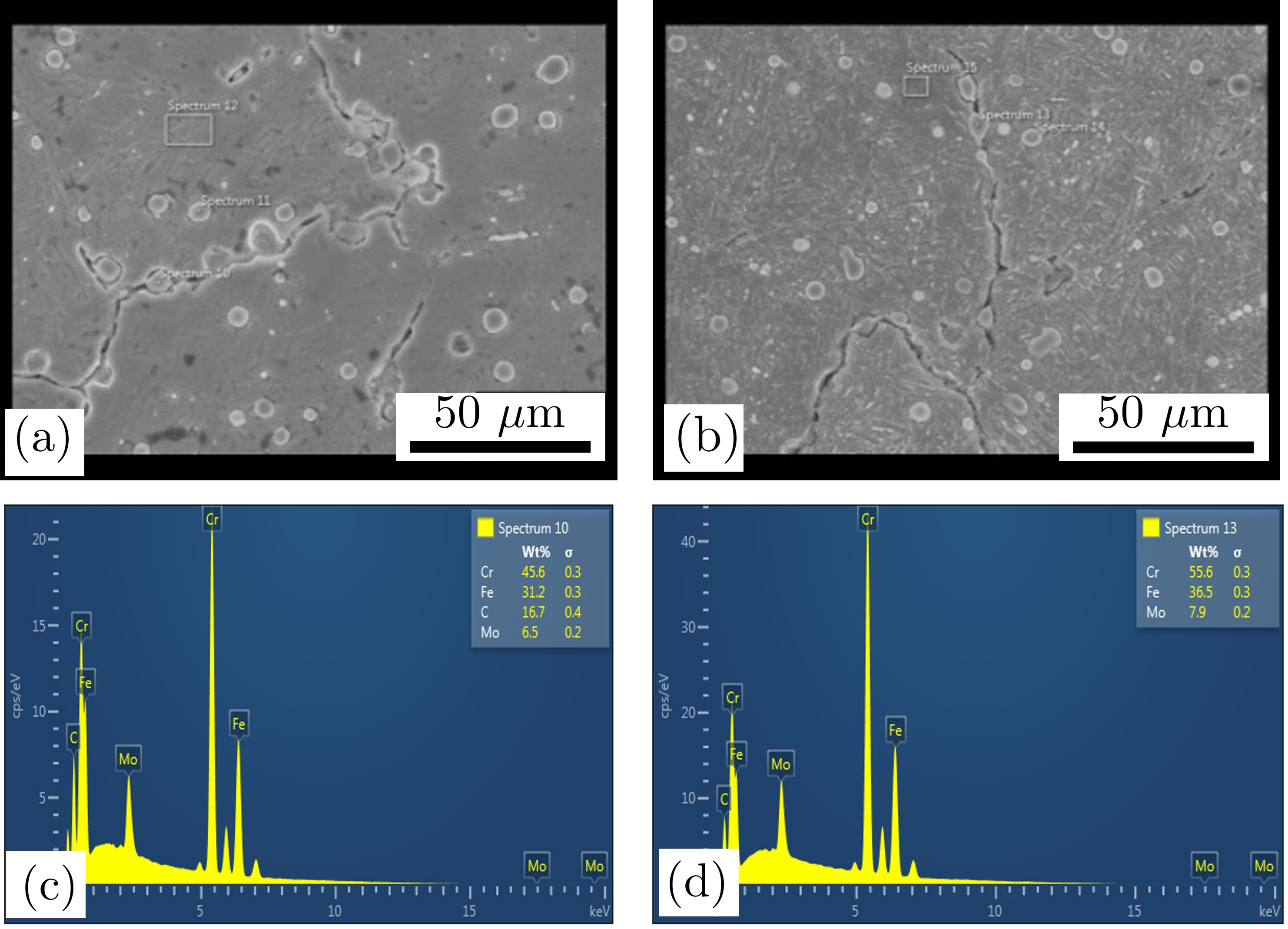}
	\caption{SEM Micrograph (a),(b) and corresponding EDS spectra (c),(d) of tempered and etched specimen at 180$^\circ$C and 525$^\circ$C respectively.}
	\label{EDS}
\end{figure*}

Precipitation of carbides and nitrides have been confirmed using X-ray diffraction analysis as well. Fig.\ref{XRD} shows the X-ray diffraction analysis of HNMS steel austenitized at 1075$^\circ$C and tempered at various temperatures. The major peak seen at 44.329$^\circ$ might belong to both martensite and austenite phases. This peak shifted towards right and shoulder peaks appeared with increase in the tempering temperature  from 180$^\circ$C to 525$^\circ$C. The additional peaks at 525$^\circ$C were identified as iron nitrides (as shown in Fig.\ref{XRD}). Subsequently, EDS analysis has been carried out to quantify the elemental composition of these precipitates.  
Fig.\ref{EDS}) shows the SEM micrographs with EDS analysis of HNMS steel austenitized at 1075$^\circ$C followed by tempering at 180$^\circ$C and 525$^\circ$C respectively. In energy dispersive spectroscopy (EDS) spectra, prominent Fe, Cr and Mo peaks were observed conforming the composition of matrix and carbides. The quantitative analysis of peaks corresponding to carbides confirms that the carbides are $M_7C_3$ (Fe, Cr, Mo)7C3 and $M_23C_7$ type. Carbides which were formed on 180°C tempering  has less Cr and Mo content compared to those formed at 525°C tempering.

Since $Fe_3C$ is least stable compound with almost negligible enthalpy of formation \cite{krauss1971morphology}. Hence, it could be expected that when strong carbide-forming elements are present in steel in sufficient concentration their carbides would formed in preference to cementite. Nevertheless, during tempering of steel, alloy carbides do not form until the temperature ranges 400-600°C. Formation of alloy carbides requires elements to diffuse in the matrix and this can happen only if sufficient activation energy is present. Metallic elements diffuse substitutionally while carbon and nitrogen diffuse interstitially, so the diffusivity of C and N is much more. Coarsening of cementite is much rapid than alloy carbide as well \cite{bhadeshia1997martensite}.
\subsection{Hardness}
\label{sec:hardness}
Since hardness is the basic mechanical property which indicates the strength and wear resistance of the material. So this has to be evaluated with different austenitizing cycles and at particular tempering temperature. The variation in hardness with austenitizing temperature is shown in figure1. Hardness values varies from 54-60HRC. The decrease in hardness with increasing austenitizing temperature is mainly because of increase in the fraction of retained austenite which can be observed from optical microscopic images. At higher austenitizing temperature more amounts of carbides gets dissolved (from SEM analysis Fig.~\ref{SEM_precipi}) resulting in higher carbon content in the matrix that stabilized the austenite phase on cooling.
\begin{figure*}[h!]
	\centering
	\includegraphics[width=\textwidth]{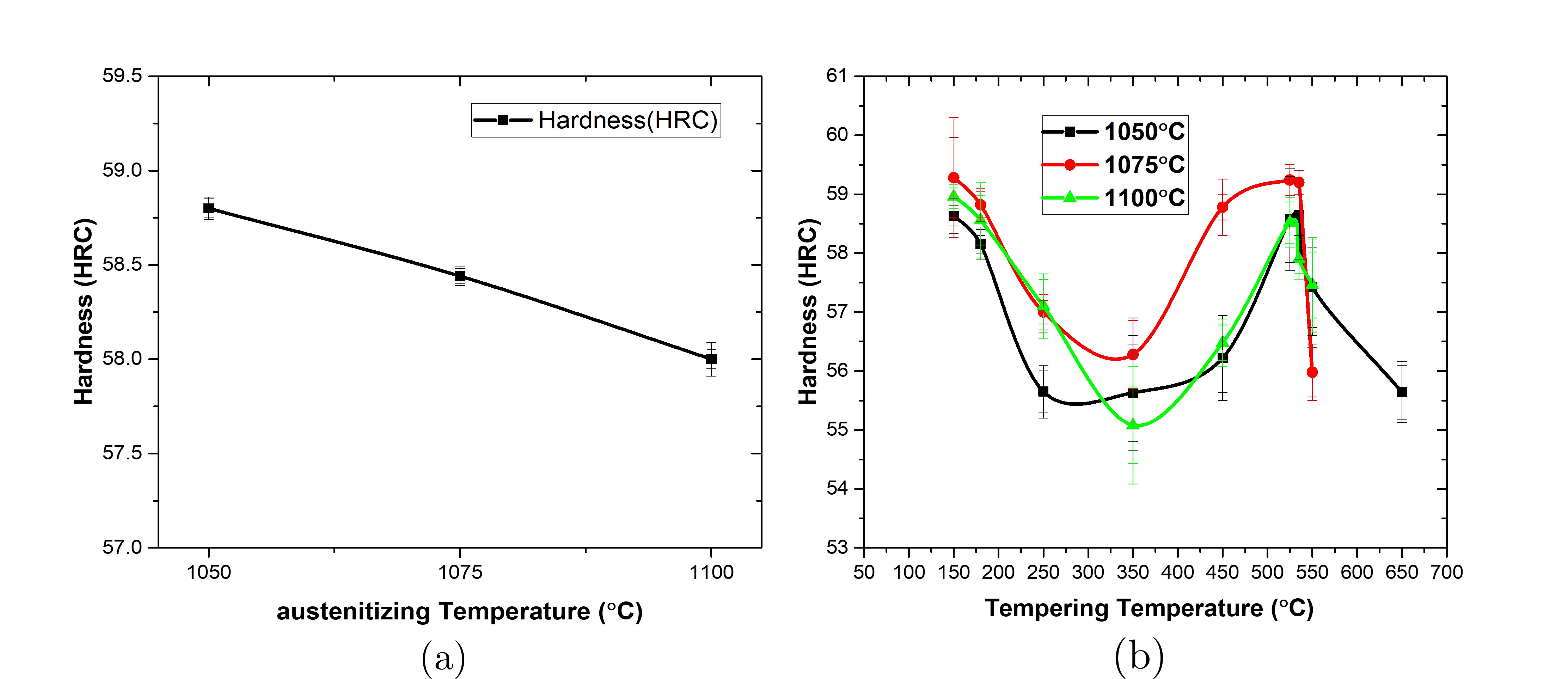}
	\caption{(a) Variation in hardness (HRC) with austenitizing at different temperatures (b) After Austenitizing at 1050 $^\circ$C, 1075$^\circ$C and 1100 $^\circ$C followed by tempering at various temperatures}
	\label{hardness}
\end{figure*}

 From Fig.~\ref{SEM_precipi} it is evident that the volume fraction of carbides in the matrix decreased with increasing austenitizing temperatures. The higher carbon content in the matrix is responsible for higher retained austenite fraction thereby reducing the hardness value. On the other hand, higher carbon content in the matrix requires a larger degree of cooling to convert all retained austenite to martensite\cite{zhu2014effect}.Fig.~\ref{hardness} (b) shows the variation of hardness with tempering temperatures for various austenitizing temperatures. The hardness values decreased with increasing tempering temperatures up to 350$^\circ$C and increased on subsequent tempering at higher temperature. The decrease in hardness at lower tempering temperature (up to 400$^\circ$C) can be attributed to the relieving of internal stresses of the quenched martensite to form tempered martensite and decrease in dislocation density. The increase in the hardness values beyond 400$^\circ$C is due to the precipitation of fine carbides and nitrides (Fig.~\ref{SEM_precipi}(a)) \cite{krishna2015effect}. The formation of secondary precipitates is responsible for the increase in the hardness values. Further increase in tempering temperature resulted in the decrease of hardness values to a minimum of 38HRC at 650$^\circ$C.  The coarsening of the precipitates results in the decrease of hardness values as shown in the Fig.~\ref{hardness}.
\begin{figure*}[h!]
	\centering
	\includegraphics[width=0.6\textwidth]{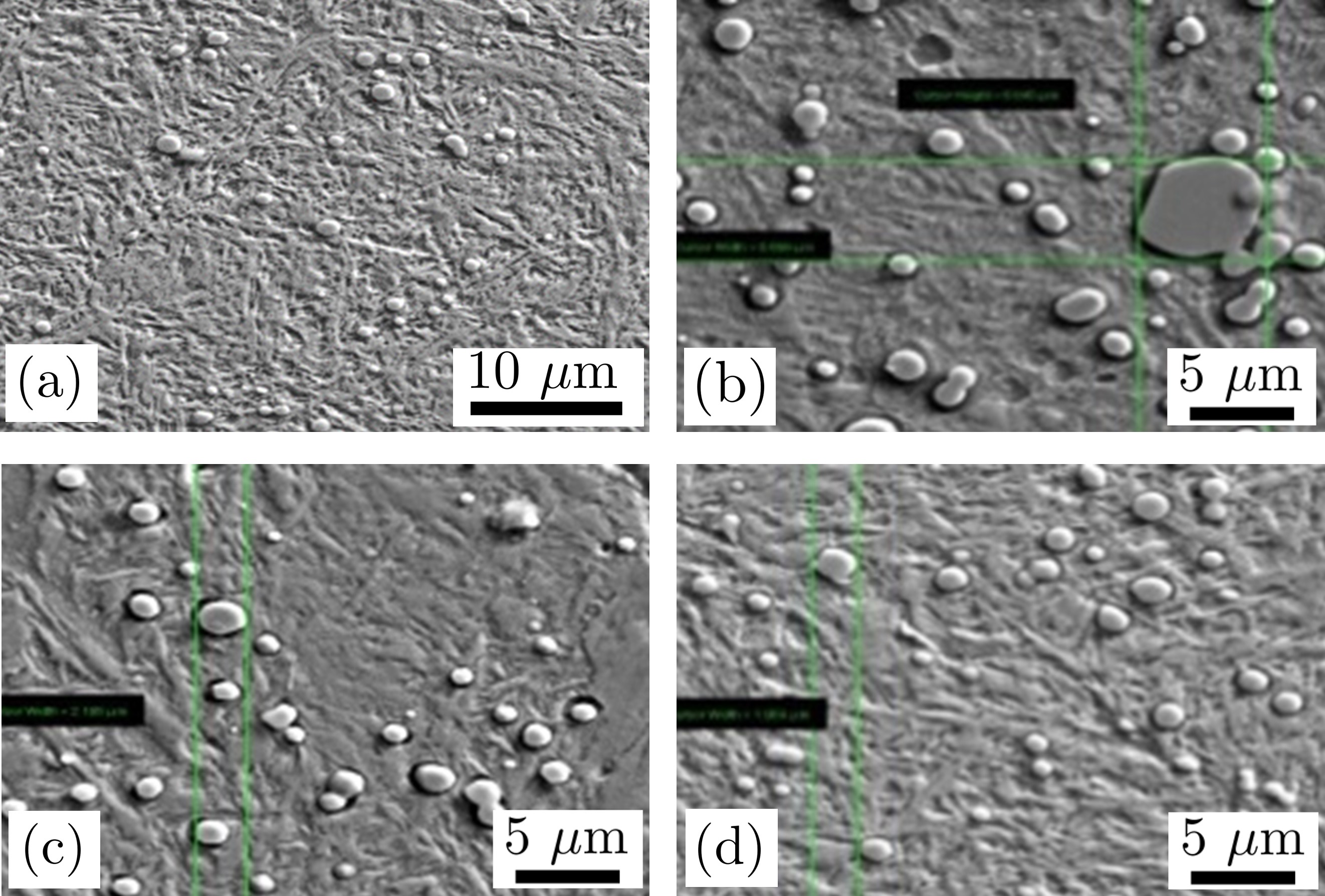}
	\caption{(a) SEM micrograph showing the precipitates after austenitizing at 1075°C followed by 525$^\circ$C tempering (b) (c) (d) After austenitizing at 1050$^\circ$C,1075$^\circ$C and 1100$^\circ$C respectively.}
	\label{SEM_precipi}
\end{figure*}

\subsection{Mechanical Testing}
Tensile test and Charpy impact test at room temperature have been carried out further to evaluate the mechanical strength of the alloy.
\subsubsection{Tensile test results}
For evaluating mechanical properties the hardening at 1075$^\circ$C followed by tempering at various temperatures was selected. Heat treatment was carried out after preparing the specimens. Tensile test was carried on 180$^\circ$C, 350$^\circ$C and 525$^\circ$C tempered specimens and results are enlisted in Table\ref{table:2}.

\begin{figure*}[h!]
	\centering
	\includegraphics[width=0.6\textwidth]{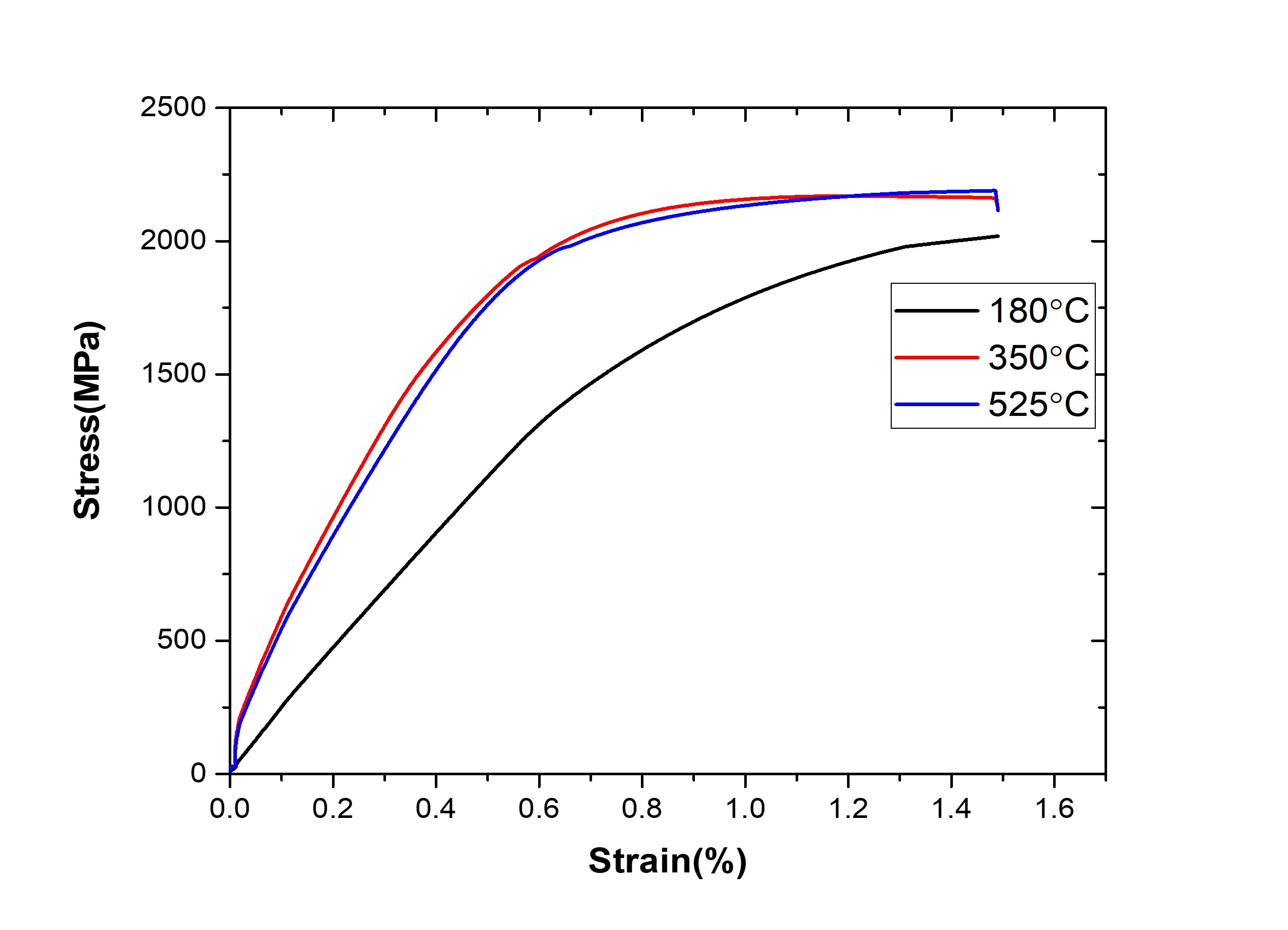}
	\caption{Stress strain curve of the sample after austenizing at 1075$^\circ$C and tempered at different temperature.}
	\label{tensile}
\end{figure*}

 It can be clearly observed that the elongation for the specimens tempered at 180$^\circ$C was observed to be almost negligible (0.5\%). The reason behind this brittle failure is mainly the residual stresses and strain present in the crystal which occurred due to the martensitic transformation. Specimens tempered at 350$^\circ$C showed a maximum elongation of 6.3\%. The decrease in the strength and maximum elongation is as a result of formation of austenite and relieving of crystal stresses \cite{lee2009reverse}. Though the elongation at 350$^\circ$C tempering is maximum but the hardness value has reduced (as discussed in Sec~\ref{sec:hardness}). Specimens tempered at 525$^\circ$C shows better Ultimate tensile strength values as well as the percentage elongation of 5.6\%. The reason behind this behavior can be attributed to the precipitation of fine $M_7C_3$ carbides and carbo-nitrides (Fig.~\ref{SEM_precipi}). Hence the specimens tempered at 525°C showed a superior tensile strength (Fig.~\ref{tensile}) along with the hardness. The results agree with the studies carried out on Fe-16Cr-1.1Mo-0.2V-0.1C-0.6N alloy\cite{ojima2008origin}. The enhanced hardness and strength were attributed to the shearing of N-enriched regions and interactions between dislocations and stress fields around these regions.
\begin{table}[ht]
	\caption{Mechanical properties after austenitizing at 1075$^\circ$C followed by tempering} 
	\centering 
	\begin{tabular}{c c c c} 
		\hline 
		Tempering Temp.($^\circ$C)& 0.2\%(MPa) & UTS(MPa) & Elongation (\%) \\ [0.5ex] 
		\hline 
		180&1700&2020 &0.5 \\ 
		350&1660&2165&6.3 \\
		525&1810&2200&5.6 \\
		\hline 
	\end{tabular}
	\label{table:2} 
\end{table}

\subsubsection{Impact test results}

Charpy impact test was carried out on specimens which were hardened at 1075$^\circ$C followed by tempering at various temperatures. The test was performed at room temperature as well as at cryogenic temperature. The values of energy absorbed are enlisted in Table\ref{table:3}.
\begin{figure*}[h!]
	\centering
	\includegraphics[width=\textwidth]{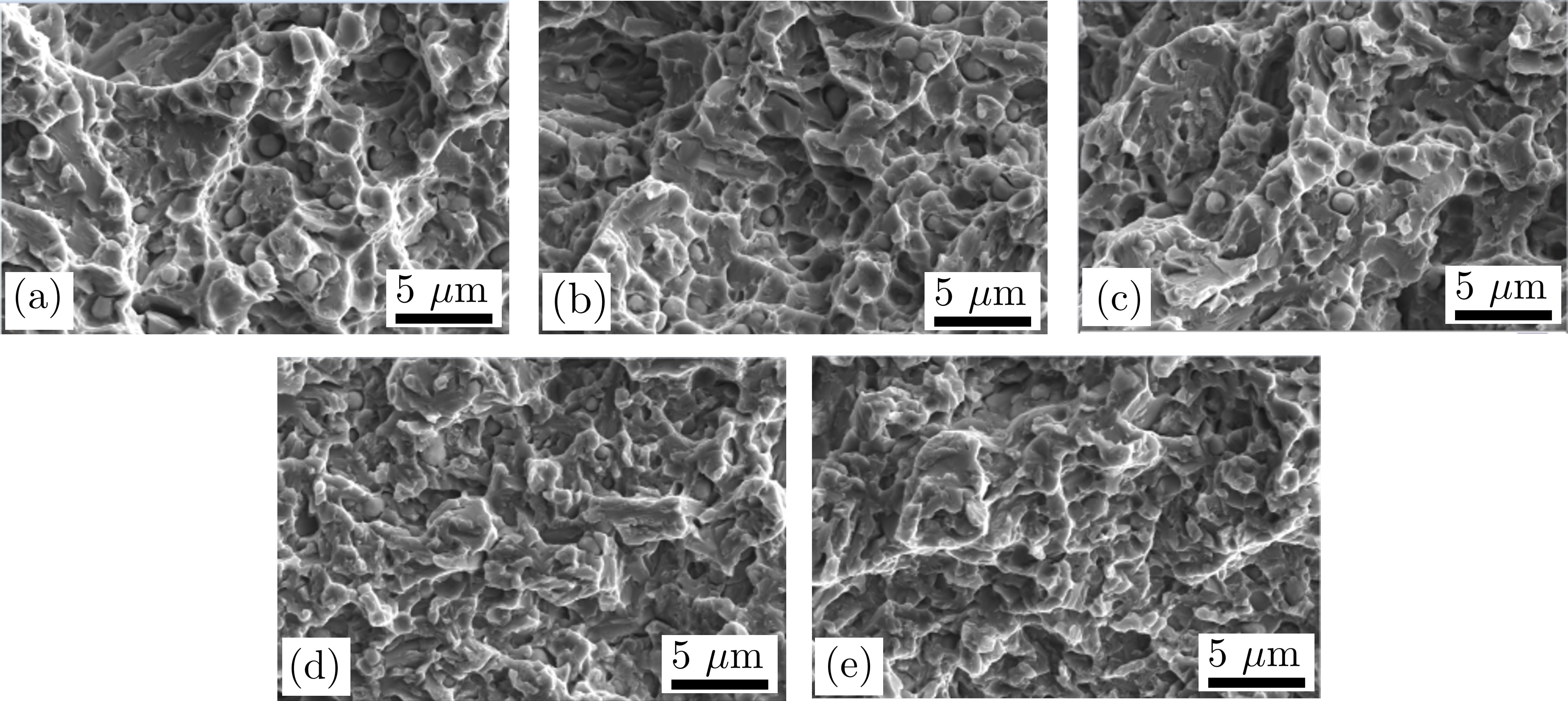}
	\caption{SEM Micrograph of fractured surface after tempering.(a),(b),(c) after tempering at 180,350 and 525$^\circ$C respectively at room temperature;(d),(e),(f) at Cryogenic Temperature.}
	\label{fracture}
\end{figure*}
. This shows the values of absorbed energy (J) with different tempering temperatures at room temperature and cryogenic temperature. Impact energy for the samples austenitized at 180$^\circ$C is minimum. This may be because of the crystal stress and distortion. On further tempering up to 350$^\circ$C the impact toughness increases but hardness decreases (as discussed in section 2) this can be attributed mainly because of the reverse austenitic transformation which is also reported in Ref.\cite{lee2009reverse}. According to which slight amount of martensite can transform into austenite on heating to relieve the crystal strain which produced on quenching. On higher temperature tempering impact toughness decreases which can be attributed to the precipitation of fine $M_7C_3$ secondary carbides and carbo-nitrides along the grain boundaries (Fig.~\ref{EDS}). 
\begin{table}[ht]
	\caption{Absorbed Impact energy of HNMS steel austenitized at 1075$^\circ$C and tempered} 
	\centering 
	\begin{tabular}{c c c c} 
		\hline 
		Tempering Temp.($^\circ$C)& 180 & 350 & 525 \\ [0.5ex] 
		\hline 
		Absorbed Energy (J) at RT&4&6 &4 \\ 
		Absorbed Energy (J) at CT&1.5&2&2 \\
		\hline 
	\end{tabular}
	\label{table:3} 
\end{table}

Nitrogen present in the steel inhibit the austenite grain growth which is also called nitride pinning effect resulted in grain refinement strengthening but lowers the impact toughness\cite{fan2012effects} . 

Fig.~\ref{fracture} shows the electron images of fractured surface after 1075$^\circ$C hardened and tempered specimens, which shows the cleavage morphology at room as well as cryogenic temperature.

\section{Conclusion}
Maximum Hardness observed was 58HRC at 1075$^\circ$C austenitized and 180$^\circ$C tempered. There is a gradual decrease in hardness with increasing tempering temperature and observed minimum at 350$^\circ$C tempered which is mainly because of the relieve in crystal stress and decrease in dislocation density. In addition to this following conclusions can be drawn from the results:
\begin{itemize}
	\item There is decrease in hardness with increasing austenitizing temperature, which is mainly because of increases in retained austenite fraction as with increase in solutionizing temperature more amount of carbides got dissolved into the matrix.

	\item The microstructure after each hardening cycle comprises mainly of martensite, carbides ($M_23C_7$, $M_7C_3$ and MC) carbonitrides (MN and M2N) and retained austenite. From FESEM with EDS and XRD analysis the phases and their composition were quantified.
	
	\item Tensile test after hardening at 1075$^\circ$C followed by tempering at 180,350 and 525$^\circ$C revealed the UTS is maximum for 525$^\circ$C Tempered and maximum elongation at 350$^\circ$C tempered. In all the cases brittle fracture were observed with negligible elongation at 180$^\circ$C tempered.

	\item Tempering temperature is the important factor that affects the hardness and mechanical properties. 1075$^\circ$C austenitized followed by 525$^\circ$C tempered condition exhibit a good combination of hardness, strength and impact toughness along with a considerable percentage elongation.
\end{itemize}

\section{Acknowledgment}

HSD would like to thank his colleagues at Aerospace Materials Laboratory (AML), LPSC/ISRO for their support in the characterization of the samples. 

\bibliographystyle{vancouver}
\bibliography{bibfile}

\end{document}